\documentclass[runningheads,citeauthoryear]{apinv2021}
\usepackage{epsfig,cite,graphics}
\usepackage{graphicx}
\usepackage{caption} 
\usepackage{marvosym} 
\usepackage[utf8]{inputenc}

\begin{document}

\title{UBVRI observations of the flickering of the dwarf nova RX~And }
\titlerunning{ Flickering of the dwarf nova RX~And  }
\author{R. K. Zamanov$^1$, G. Nikolov$^{1,2}$, A. T. Georgieva$^1$}
\authorrunning{Zamanov,  Nikolov \& Georgieva}
\tocauthor{R. K. Zamanov, G. Nikolov,  A. T. Georgieva} 
\institute{Institute of Astronomy and National Astronomical Observatory, Bulgarian Academy of Sciences, Tsarigradsko Shose 72, 
BG-1784 Sofia, Bulgaria  \\
	\and  Westinghouse Electric, Business Park Sofia,  building 8A, floor 7,  1766 Sofia, Bulgaria  \\
	\email{rkz@astro.bas.bg  \hskip 0.3cm  gnikolov@astro.bas.bg  \hskip 0.3cm georgieva.a.t@gmail.com}      }
\papertype{Submitted on 26.03.2021; Accepted on 11.04.2021}	
\maketitle

\begin{abstract}
We report observations of the flickering variability of the  dwarf nova  RX And
in five bands (UBVRI) on two nights. 
On 25 October 2019 the brightness of the star 
was $B\approx 13.9$~mag, the amplitude of the flickering was 0.47 mag, and we estimate for the flickering source 
temperature $T_{fl} = 10700 \pm 400$~K, and radius $R_{fl} =0.046 \pm 0.004$~$R_\odot$.
On  2 January 2020, the star was about 3 magnitudes brighter 
($B \approx 10.7$), the amplitude of the flickering was significantly lower (0.07 mag) and we derive for the flickering source 
$T_{fl} = 9600 \pm 700$~K, and radius $R_{fl} = 0.098 \pm 0.009$~$R_\odot$.
The results indicate that 3 magnitudes brightening of the star
doubled the radius of the flickering source. \\ 
The data are available upon request from the authors.
\end{abstract}

\section{Introduction}

RX~And (2MASS J01043553+4117577) is a  cataclysmic variable  of dwarf nova type
with orbital period $0.2098930$ day = 5h02m = 302 minutes  (Kaitchuck 1989). 
The cataclysmic variables are short period binaries consisting 
of an white dwarf primary and a red dwarf as a mass donor.
A sketch representing a cataclysmic variable binary is included in the Appendix. 

The underlying white dwarf in RX And was first detected 
by Holm et al. (1991) using an IUE spectrum
during RX And's quiescence.
The infrared spectra identified that the mass donor is a K5V star 
(Dhillon \&  Marsh 1995). 


The dwarf novae are cataclysmic variables which exhibit
recurrent outbursts with amplitude of 2 to 5 mag on the  time-scale of weeks, 
caused by an increase in the mass accretion rate. 
RX And  shows regular dwarf nova outbursts  every 15 days (e.g. Kato 2004)
and flickering on time scale of minutes (e.g. Bennert et al. 1999).

Following the AAVSO light curve generator, during the last 
years RX And varies in the range $10.9 \le V \le 14.9$ and
$9.5 \le B \le 15.0$. There was an extended faint state of RX And in 1996, 
when the brightness remained about 15 mag in V band for $\approx 100$~days 
(Kato et al. 2002).

In this work, we present  quasi-simultaneous 
UBVRI observations of the flickering variability of RX~And
and estimate the parameters of the flickering source.

\section{Observations}

The observations are performed  with the 50/70 cm Schmidt telescope of the 
National Astronomical Observatory Rozhen repeating U,B,V,R and I bands --
on   25 October 2019 from UT~18:33 till 00:20, total duration  5h~47m, 
and on 2 January 2020 from UT~17:43 till 21:31, total duration 3h 48m. 
The telescope was equipped with a CCD camera
and the field of view was 23 x 23 arcmin.  
The red bands correspond to the Cousins VRI system.

The comparison stars were  TYC~2807-1285-1 (U=11.21, B=11.15, V=10.57, R=9.88, I=9.53)
and
TYC~2803-1045-1  (U=11.96, B=11.94, V=11.43, R= 10.75, I=10.44).
The data reduction is done with IRAF (Tody 1993) 
following the standard recipes for processing of CCD images and aperture photometry. 

In  Table~\ref{tab2} are given date of observation, its duration in minutes, 
number of the exposures ($N_{pts}$), the exposure times in seconds, 
minimum,  maximum and average magnitude in the corresponding band,  
standard deviation of the run, 
typical observational error 
and peak-to-peak amplitude of the variability. 

\begin{figure}    
   \vspace{8.7cm}     
   \includegraphics{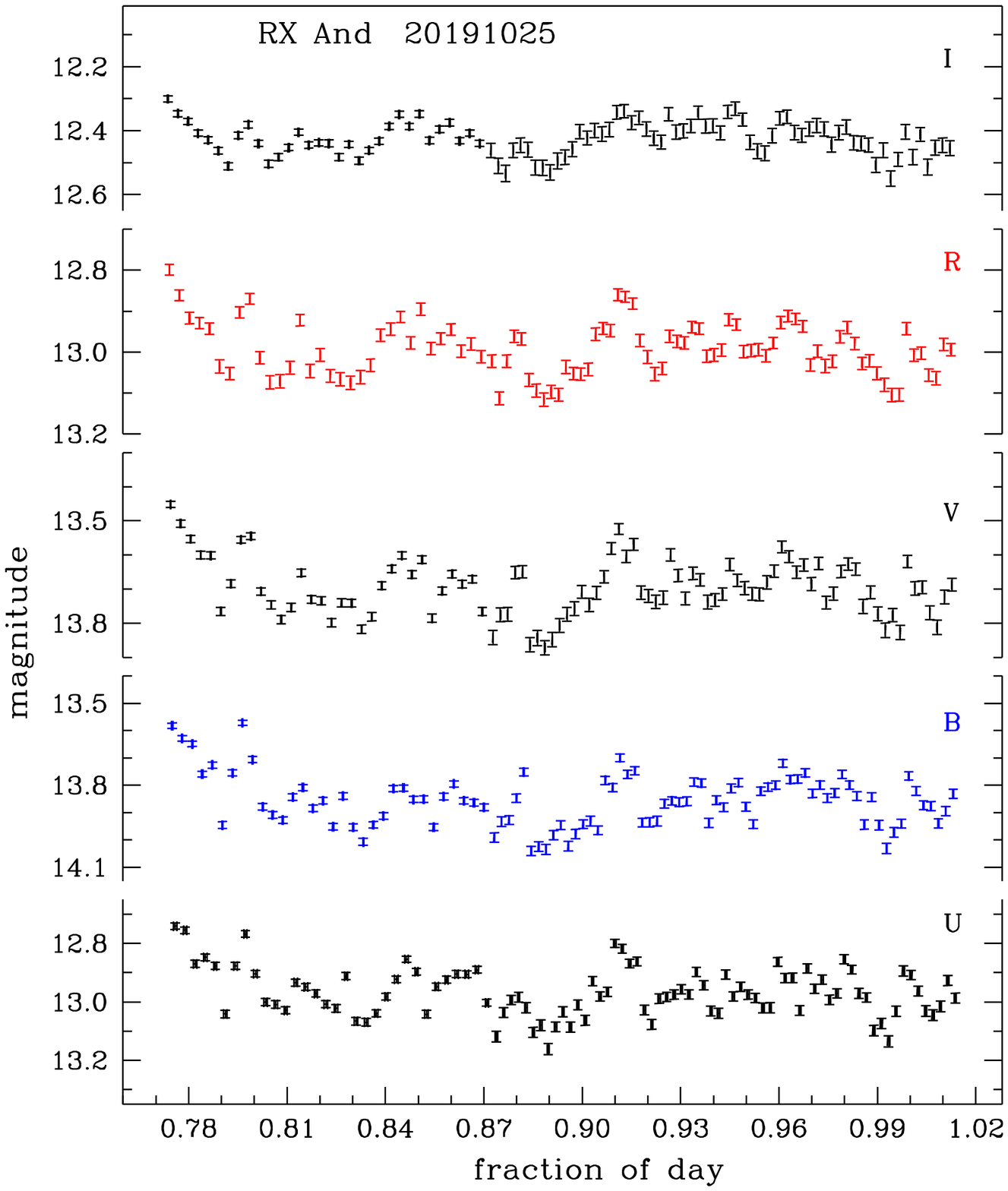}  
   \includegraphics{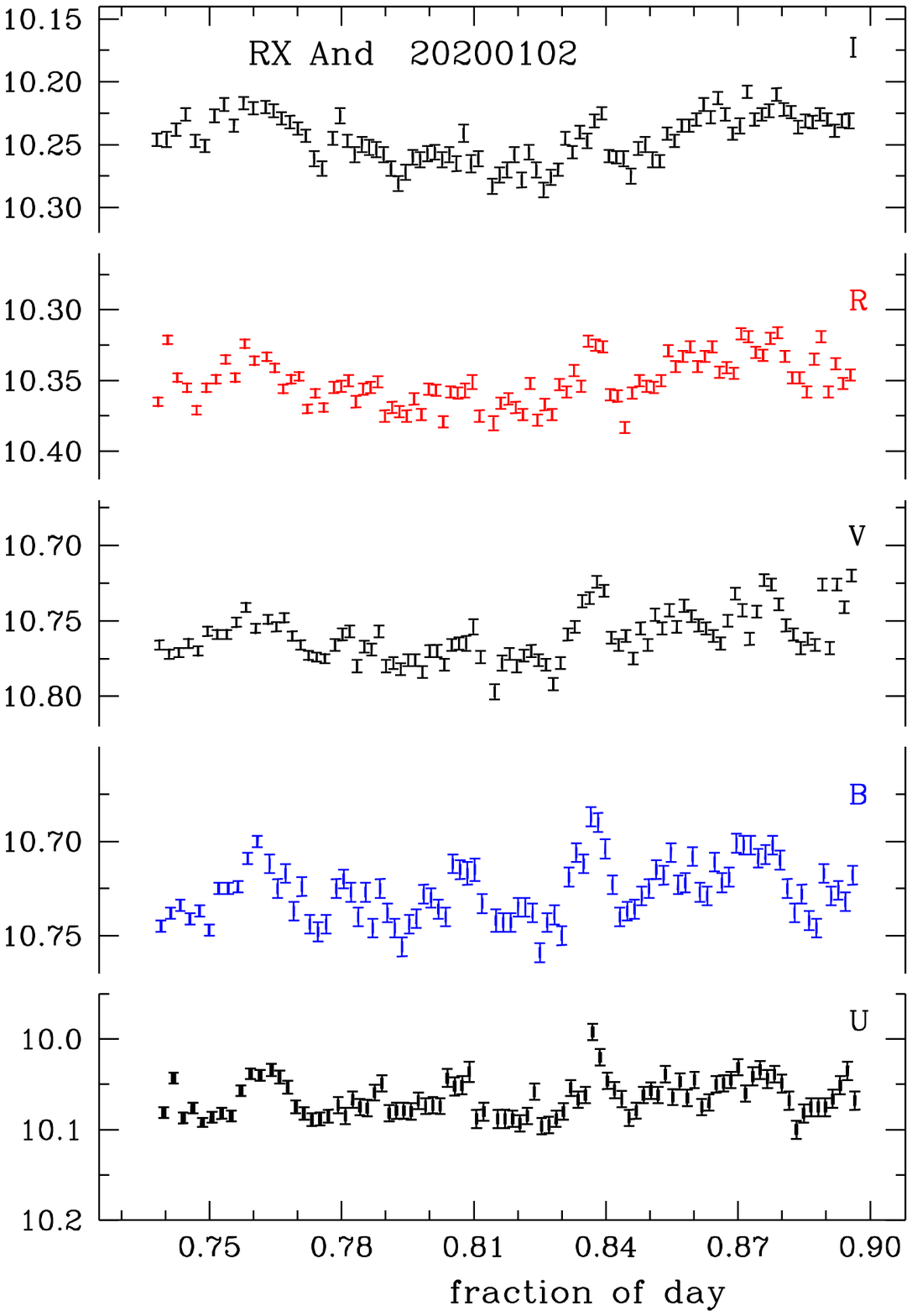}  
   \caption[]{UBVRI observations of the flickering dwarf nova RX~And on 25 October 2019 and 2 January 2020
   performed with the 50/70~cm Schmidt telescope of NAO Rozhen.}
   \label{f.obs} 
 \end{figure}

Our observations are presented on Fig.~\ref{f.obs}. 
What is immediately visible is that
during  the second night the star was brighter and the amplitude of the flickering was considerably lower.  


  \begin{table}
  \begin{center}
  \caption{Photometry of RX And.  In the table are given: date of observation, its duration, 
  band, number of the data points obtained, 
  exposure time [in seconds],  
  minimum, maximum and average magnitudes  in the corresponding band, standard
  deviation of the mean, typical observational error, peak-to-peak amplitude.
  }
  \begin{tabular}{l c l  r   c  | c   c c c  c r r rrr}
 date        &  band & $N_{pts}$ & exptime &    &   &  min    &  max   & average  & stdev  &  merr  & ampl. \\
 duration    &       &           &  [sec]  &    &   & [mag]   & [mag]  &  [mag]   &  [mag] &  [mag] & [mag] \\
             &       &     &		   &    &   &	      &        &	  &	   &	    &	    \\  	  
 \hline
             &       &     &	           &    &   &	 &	  &	     &        &        &       \\		      
 2019-10-25  &  U    & 95  &  90, 60       &    &   &  12.741 & 13.161 &  12.9680  &  0.084 & 0.017  &  0.42 \\
   347 min   &  B    & 95  &  30, 20       &    &   &  13.571 & 14.040 &  13.8571  &  0.097 & 0.017  &  0.47 \\
             &  V    & 95  &  20, 10       &    &   &  13.452 & 13.871 &  13.6971  &  0.084 & 0.020  &  0.42 \\
             &  R    & 95  &  10           &    &   &  12.799 & 13.116 &  12.9934  &  0.065 & 0.015  &  0.32 \\
	     &  I    & 95  &  30, 10       &    &   &  12.301 & 12.550 &  12.4255  &  0.053 & 0.023  &  0.25 \\
             &       &     &	           &    &   &	     &        & 	  &	   &	    &	    \\
 2020-01-02  &  U    & 90  &  60, 40, 30   &    &   &	9.992 & 10.100 &  10.0652  &  0.020 & 0.009  &  0.11 \\
  228 min    &  B    & 90  &  20, 10       &    &   &  10.687 & 10.759 &  10.7267  &  0.015 & 0.006  &  0.07 \\
	     &  V    & 90  &  10,  7       &    &   &  10.720 & 10.797 &  10.7598  &  0.016 & 0.005  &  0.08 \\
	     &  R    & 90  &  10,  7       &    &   &  10.316 & 10.383 &  10.3504  &  0.017 & 0.005  &  0.07 \\
	     &  I    & 90  &  10           &    &   &  10.208 & 10.286 &  10.2454  &  0.019 & 0.006  &  0.08 \\
             &       &     &	           &    &   &	     &        & 	 &	  &	   &	   \\
 \\
 \label{tab2}
 \end{tabular}
 \end{center}
  \begin{center}
  \caption{Estimated colours of RX~And and its flickering source.
   The colours of the star are the average colours during the night. 
   }
  \begin{tabular}{c c c |  r  r   c | r r  r c r r}
  
 date    &    &  colour & &   &  star		 &   &  average     &	   maximum     &  \\
         &    & 	& &   & 		 &   &  flickering  &	   flickering  &  \\
 \hline
         &    & 	& &   & 		 &   &  	      & 	     & \\		   
20191025 &    &  (U-B)  & &   & $ 0.89\pm0.02$   &   & $-0.94\pm0.03$ & $-0.73\pm0.05$ & \\
         &    &  (B-V)  & &   & $ 0.16\pm0.02$   &   & $ 0.11\pm0.03$ & $ 0.02\pm0.05$ & \\
         &    &  (V-R)  & &   & $ 0.70\pm0.02$   &   & $ 0.35\pm0.04$ & $ 0.40\pm0.07$ & \\
         &    &  (R-I)  & &   & $ 0.57\pm0.02$   &   & $ 0.58\pm0.06$ & $ 0.27\pm0.09$ & \\
  &    &   &    &   &    &   &    & \\
20200102 &    &  (U-B)  & &   & $ 0.66\pm0.01$   &   & $-0.74\pm0.09$ & $-1.12\pm0.09$ & \\ 
         &    &  (B-V)  & &   & $-0.03\pm0.01$   &   & $ 0.12\pm0.05$ & $ 0.04\pm0.05$ & \\
         &    &  (V-R)  & &   & $ 0.41\pm0.01$   &   & $ 0.27\pm0.04$ & $ 0.26\pm0.04$ & \\
         &    &  (R-I)  & &   & $ 0.11\pm0.01$   &   & $ 0.34\pm0.03$ & $ 0.27\pm0.04$ & \\
         &    & 	& &   & 		 &   &  	      & 	     & \\
 \hline 
 \\
 \label{tab3}
 \end{tabular}
 \end{center} 
\end{table}

\section{Flickering light source}
\label{s.fli}

Bruch (1992) proposed that the light curve of a white dwarf with flickering  can be separated into two parts -- constant light
and variable (flickering) source. 
Following his recipe, 
we calculate the flux of the flickering light source 
as $F_{\rm fl}=F_{\rm av}-F_{\rm min}$, where $F_{\rm av}$ is the average flux 
during the run and $F_{\rm min}$ is the minimum flux during the run
(corrected for the typical error of the observations).
An expansion of the method is proposed by
Nelson et al. (2011). 
They suggest to use the $F_{\rm fl.max}=F_{\rm max}-F_{\rm min}$, where $F_{\rm max}$ 
is the maximum flux during the run. 
In fact, the method of Bruch (1992) evaluates the average brightness of the flickering source, 
while that of Nelson et al. (2011) --  its maximal brightness. 
$F_{\rm fl}$ and $F_{\rm fl.max}$  have been calculated for each band, using the values 
given in Table~\ref{tab2} and 
the calibration for a zero magnitude star (Bessell 1979). 
  
Following the IRSA: Galactic Reddening and Extinction Calculator
in NASA/IPAC Extragalactic Database, NED 
(which  is operated by the Jet Propulsion Laboratory,  California Institute of Technology), 
the extinction toward RX~And is low, $E(B-V) \le 0.06$. 
The IRSA calculator uses Galactic reddening maps to determine the total Galactic line-of-sight reddening
(in front of RX And and behind it),
and is based on the results by  Schlegel, Finkbeiner \& Davis (1998)  
and Schlafly \& Finkbeiner (2011).

GAIA EDR3 (Lindegren et al. 2021) gives  parallax $5.0510 \pm 0.0269$~mas.
Hereafter, we assume no extinction, $E(B-V) = 0.0$ and distance $198 \pm 1$~pc. 
We calculate for the average flickering source of RX~And 
its temperature and radius: \\
20191025 -- \hskip 0.1cm  $T_{fl} = 10700 \pm 400$~K,  $R_{fl} = 0.046 \pm 0.004$~R$_\odot$,   \\
20200102 -- \hskip 0.3cm  $T_{fl} =  9600 \pm 700$~K,  $R_{fl} = 0.098 \pm 0.009$~R$_\odot$.   \\
These values correspond to luminosity $L_{fl}=0.025\pm0.004$~L$_\odot$ (for 20191025),
and $L_{fl}=0.074\pm0.008$~L$_\odot$ (for 20200102). 


 \begin{figure}    
   \vspace{7.0cm}     
   \includegraphics{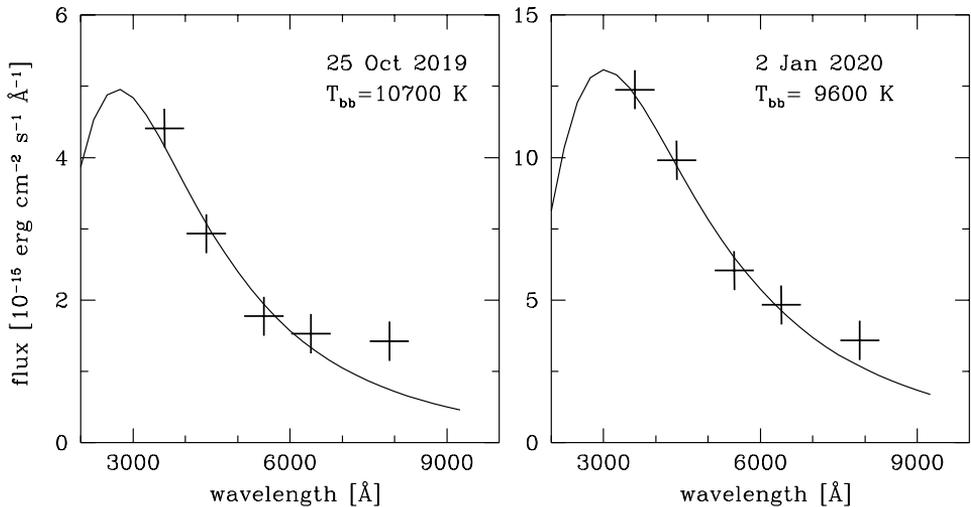}  
   \caption[]{Black body fits to the flickering source of RX~And.}
   \label{f.bb} 
 \end{figure}        

\section{Discussion}
\label{D.1} 
The light curve of RX And  in 1972-1997 displays periods
of frequent outbursts with amplitude $\approx 3$~mag, low states and periods of stand still
(Schreiber et al. 2002). 
The light curve of RX And in  2014-2015 was examined by Timar (2016), 
who found that the cycle of outbursts is short ($< 15$ days) due to the high accretion rate
and every fourth outburst is a super-outburst. The period of the super-cycle was about 55.5 days.

The Hubble Space Telescope spectra revealed that the white dwarf in RX~And has a
temperature $T_{wd} \approx 34000$~K,   
and  rotates with $\approx 600$~km~s$^{-1}$ (Sion et al. 2001). 
Sepinsky et al. (2002) using HST observations  during outburst rise and decline 
found  mass accretion rate $2 \times 10^{-10}$ M$_\odot$~yr$^{-1}$ onto 
an white dwarf of mass $\sim 0.8$~M$_\odot$ with effective temperature 40000~K.
Godon \& Sion (2003) found that it must be a massive white dwarf $\sim  1.2$~M$_\odot$. 


The temperature of the flickering source estimated here 
is considerably lower than the white dwarf temperature. 
It is  similar to 
the temperature of the bright spot of the cataclysmic variables. 
For the bright spot of  OY Car  Wood et al (1989) calculated
black body temperature $T = 13800 \pm 1300$ K, and color temperature $T = 9000$~K.
Marsh (1988) for IP Peg give $T = 11200$ K, Zhang \& Robinson (1987) -- for U Gem - $T = 11600 \pm 500$~K, 
Robinson et al. (1978) -- $T = 16000$~K for the bright spot in WZ Sge. 
The colours of the flickering source of RX And 
on 10 August 1983 were estimated by Bruch (1992)  $U-B=-1.14$,  $B-V=0.46$, assuming E(B-V)=0.06.
The value of U-B is bluer than our and B-V is more red. The differences 
can be connected with real changes in the flickering and/or with the used different calibrations 
for a zero magnitude star.

In dwarf novae, the flickering amplitude is high during quiescence, 
drops quickly at an intermediate magnitude when the system enters into (or returns from) an outburst and, 
on average, remains constant above a given brightness threshold (Bruch 2021). 
In future it will be interesting to search for a correlation between the flickering source parameters 
and brightness as done for the recurrent nova RS~Oph (Zamanov et al. 2018)
and the jet-ejecting symbiotic MWC~560 (Zamanov et al. 2020).

\vskip 0.1cm 

{\bf Conclusions: }
We report quasi-simultaneous observations in 5 bands (UBVRI) of the flickering 
of the cataclysmic variable RX And during two nights.

For 25 October 2019 the brightness of the star 
was $ 13.57 \le  B \le 14.04$, the amplitude of the flickering in B band was $\Delta B = 0.47$ mag. 
For the optical flickering source we obtained colour $(B-V)_0 = 0.11$, 
temperature $T_{fl} = 10700 \pm 400$~K, and radius $R_{fl} =0.046$~$R_\odot$.  

For  2 January 2020, the star was about 3 magnitudes brighter in B band
$ 10.69 \le B \le 10.76$
and the amplitude of the flickering decreased considerably  $\Delta B = 0.07$ mag.
For the optical flickering source we derived 
colour $(B-V)_0 = 0.12$, 
temperature $T_{fl} = 9600 \pm 700$~K, and radius $R_{fl} = 0.098$~$R_\odot$.  

The results indicate that when the star is 3 magnitudes brighter
the radius of the flickering source is twice larger. 

\vskip 0.1cm 

{\small {\bf Acknowledgments: }
This work was supported by the  Bulgarian National Science Fund project  number K$\Pi$-06-H28/2 08.12.2018
"Binary stars with compact object".
We are grateful to the referee  prof. N. A. Tomov for useful comments.



\vskip 0.1cm 

{\bf Appendix}
 \begin{figure}    
   \vspace{7.7cm}     
   \includegraphics{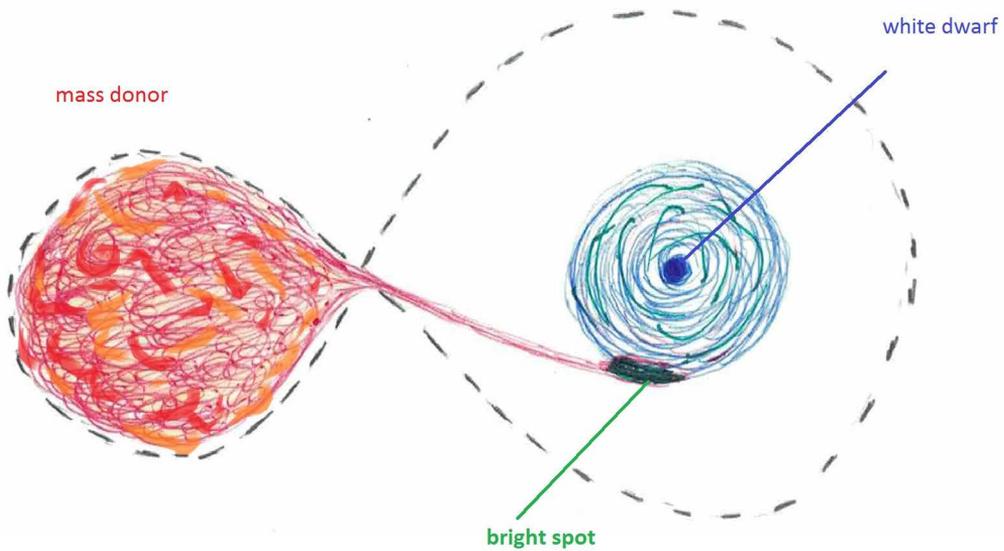}  
   \caption[]{A sketch representing the dwarf nova RX And. The mass donor is a K5V star filling its Roche lobe. 
   A stream of matter from the mass donor is feeding the accretion disc around the white dwarf. }
   \label{f.ske} 
 \end{figure}      

\begin{thebibliography}{}
\bibitem{} Bennert , N.,  Konig, I.,  Manthey, E., et al., 1999, IBVS 4779, 1
\bibitem{} Bessell, M. S. 1979, PASP, 91, 589 
\bibitem{} Bruch, A., 1992, A\&A, 266, 237
\bibitem{} Bruch, A., 2021, MNRAS, 503, 953 
\bibitem{} Dhillon, V. S. \&  Marsh, T. R., 1995, MNRAS,  275, 89
\bibitem{} Godon P. \& Sion  E. M.,  2003, ApJ, 586,  427
\bibitem{} Holm, A., Lanning, H., Mattei, J. A., Nelan, E. 1991, JAAVSO, 20, 166
\bibitem{} Kaitchuck, R. H., 1989, PASP, 101,  1129
\bibitem{} Kato, T., Nogami, D., Masuda, S.,  2002, PASJ, 54, 575 
\bibitem{} Kato, T., 2004, PASJ, 56, S55
\bibitem{} Lindegren, L., Klioner, S. A.,  Hernandez, J., et al., 2021, A\&A, in press   (arXiv:2012.03380)
\bibitem{} Marsh, T. R. 1988, MNRAS, 231, 1117
\bibitem{} Nelson, T., Mukai, K., Orio, M., Luna, G.~J.~M., Sokoloski, J.~L. 2011, ApJ, 737, 7
\bibitem{} Robinson, E. L., Nather, R. E., \& Patterson, J. 1978, ApJ, 219, 168
\bibitem{} Schlafly, E. F.,  Finkbeiner, D. P., 2011, \apj, 737, 103  
\bibitem{} Schlegel, D. J, Finkbeiner, D. P., Davis,M., 1998, \apj, 500, 525   
\bibitem{} Schreiber, M. R.,  Gansicke, B.~T.,  Mattei, J.~A., 2002, ASP Conf., 261, 545 
\bibitem{} Sepinsky, J. F., Sion, E. M., Szkody, P., Gansicke, B. T., 2002, ApJ, 574, 937
\bibitem{} Sion, E. M., Szkody, P., Gaensicke, B., Cheng, F. H.,  La Dous, C., Hassall, B., 2001, ApJ, 555, 834
\bibitem{} Timar, A., 2016, Journal of the American Association of Variable Star Observers, 44, 3
\bibitem{} Tody, D. 1993, ASP Conf., 52, 173.
\bibitem{} Wood, J. H., Horne, K., Berriman, G., \& Wade, R. A. 1989, ApJ,  341, 974
\bibitem{} Zamanov, R. K., Boeva, S., Latev, G. Y., Marti, J., Boneva, D.,  Spassov, B.,  Nikolov, Y.,  
            Bode, M. F.,  Tsvetkova, S. V.,  Stoyanov, K. A.,  2018, MNRAS, 480, 1363 
\bibitem{} Zamanov, R. K.,  Boeva, S.,  Stoyanov, K. A.,  Latev, G., Spassov, B., Kurtenkov, A., Nikolov, G., 2020, AN, 341, 430
\bibitem{} Zhang, E.-H., \& Robinson, E. L. 1987, ApJ, 321, 813


\end{thebibliography}
\end{document}